# Determining Linker Ratios of Mixed Metal-Organic Frameworks via Magnetic Susceptibility Measurements

Na Du,[‡1] Miao Miao Zhao,[‡1] Xintian Wang,[‡1] Yan Zhao,[‡1] Yang Yang,[1] Yu Ying Zhu,[2] Ruo Tong Wang,[2] Peng Ren,*[1] Fei Yen*[1]

[1]School of Science, Harbin Institute of Technology, Shenzhen, University Town, Shenzhen, Guangdong 518055, P. R. China.

[2]School of Mechanical Engineering and Automation, Harbin Institute of Technology, Shenzhen, University Town, Shenzhen, Guangdong 518055, P. R. China.

ABSTRACT

Partial replacement of the organic linkers of metal-organic frameworks (MOFs) often optimizes their functionalities, however, accurate characterization of their molar ratios in many cases is challenging. This work presents a method of determining such linker ratios via measurements of the magnetic susceptibility of small quantities of powdered samples. The main presumption is taking the diamagnetic and paramagnetic contributions to the *molar* magnetic susceptibility of the two parent MOFs to be additive. To verify this, four examples are provided with commonly used MOFs to represent the cases when both parent MOFs are either paramagnetic or diamagnetic but with different linkers with the following systems: [MIL-101(Cr)-SO$_3$H]$_{(1-\delta)}$[MIL-101(Cr)-NO$_2$]$_\delta$, [EuMOF]$_{(1-\delta)}$[EuPDCA]$_\delta$, [UiO-66-COOH]$_{(1-\delta)}$[UiO-66]$_\delta$ and [MIL-101(Cr) F Free]$_{(1-\delta)}$[MIL-53(Al)]$_\delta$, where $1-\delta : \delta$ are the ratios to be determined. Depending on whether the systems were strictly paramagnetic, strictly diamagnetic or mixed, the experimental error of $\delta$ ranged between 0.00002 and 0.012, respectively. We expect the presented method to be widely employed since samples only need to be in powdered form and because there is a lack of characterization tools in the area of MOF linker ratios. The presented method is also applicable to resolving the ratios of mixed ordinary paramagnetic systems as well as other types of non-magnetic composite materials such as tapes, zeolites and thin films.

INTRODUCTION

In metal-organic frameworks (MOFs), partial replacement of a particular metal with another node or a ligand by another linker is common as this approach often enhances the functionalities of gas adsorption, catalysis, biosensing and/or drug delivery[1-5] as well as fine-tune the stabilities of these materials.[6-9] As such, knowledge of the stoichiometric ratio of the different metals or linkers within an MOF is of vital importance.[6,10,11] However, tools often employed to determine doping concentrations of mixed solids are not applicable or may not be accurate enough for the case of MOFs. For instance, X-ray diffraction patterns of mixed MOFs with the same crystal structure as its parent compounds are often indistinguishable so working with lattice constants is rather challenging even if single crystals can be grown. Solution NMR spectroscopy can determine linker ratios but it requires a large amount of sample that needs to be dissolved, usually >1 cm$^3$.[6,11] Inductively coupled plasma mass spectroscopy (ICP-MS) can perform elemental analysis to great accuracy (~ 1 ppm); however, only metal node ratios can be discerned while partial substitution of linkers are not easy to resolve when both parent compounds do not have a unique element to serve as a tracer. Other methods used to determine linker ratios are also available but they are more elaborate and are limited to only a small subset of MOFs: for example, Fourier transform infrared spectroscopy is useful when the functional groups are infrared-active[10] and certain mixed MOFs acting as the stationary phase in liquid chromatography exhibit systematic levels of retention of the mobile phase.[12] As such, there seems to be a void in the arsenal of MOF characterization tools in the region of accurate determination of linker ratios.

In an earlier article we presented a method to obtaining the chemical composition of a mixed solid comprised of two parent compounds A and B via measurements of the magnetic susceptibility.[7] For the general formula $A_{1-\delta}B_{\delta}$, where $\delta$ is the doping concentration, $\delta$ can be determined by only knowing the molar masses (molecular weight of the unit cell) of A and B and measuring the mass magnetic susceptibility of A, B and that of the mixed solid. The problem with this method is that A and B have to be diamagnetic. Indeed, we tested this method on mixed solids comprised of UiO-66 and UiO-66-Br MOF parent compounds and obtained an average experimental error < 0.007 for $\delta$ with only 10.53 mg of powdered sample.[7] While this method is still applicable to many inorganic compounds and most organic compounds, many MOFs are also paramagnetic at room temperature. In some MOFs there may exist antiferromagnetic interactions within dimers or trimers,[8,9] however, adjacent units of the same species are too far away from each other to interact intermolecularly. According to Mínguez Espallargas & Coronado, cooperative magnetism typically appears "much below 100 K".[13] Hence, all non-activated MOFs to date, are either diamagnetic or can be treated to be comprised of paramagnetic sites at room temperature. These unpaired electrons have a similarity with the paired electrons residing in the core shells and bonds in that they do not experience intermolecular interactions so our previous approach on diamagnets should also be applicable to paramagnetic MOFs.

In this work, we first provide a brief introduction to the theory behind the employed method expanded to include paramagnets. Then, four examples on representative MOFs are worked out in detail as proof-of-concept. We finish with an error analysis to highlight the main advantages of the proposed method: linker ratio determination of mixed MOFs with minimal sample requirements (around 10 mg, powder pressed into pellet, no further purification needed) using rather accessible laboratory equipment (magnetometer and balance) in a small amount of time.

**Method of obtaining $\delta$ of a mixed MOF $A_{1-\delta}B_\delta$ comprised of pure A and B MOFs.**

Provided that no physical interactions or chemical reactions occur, the molar magnetic susceptibility $\chi_{Mol}$ follows the additivity rule. This means that if compounds A and B had magnetic susceptibilities $\chi_{Mol\_A}$ and $\chi_{Mol\_B}$, respectively, then the molar magnetic susceptibility of any mixture of A and B should be $\chi_{Mol\_A} \cdot (1-\delta) + \chi_{Mol\_B} \cdot \delta$, where $(1-\delta) : \delta$ is the mixing ratio of A and B. Note that in contrast to the molar magnetic susceptibility, the mass and volume magnetic susceptibilities are not additive, *i.e.* they do not behave linearly with respect to $\delta$.

For a diamagnet, $\chi_{Mol\_D} = M_{Mol\_D} / H$, where $M_{Mol}$ is the molar magnetization and $H$ the applied magnetic field. All solids have a diamagnetic component because the electrons will react to $H$ by orbital motion so as to generate an opposite magnetic field to cancel the effects of $H$, *i.e.* conservation of charge. The magnetization effects of $H$ on these core-shell and bonded electrons cause their individual magnetic moments to mostly align along the opposite direction so $\chi_{Mol\_D}$ is negative for all diamagnetic compounds. $\chi_{Mol\_D}$ can be experimentally obtained from the slope of a linear fit of the $M_{Mol\_D}$ vs. $H$ curve (red line in Figure 1).

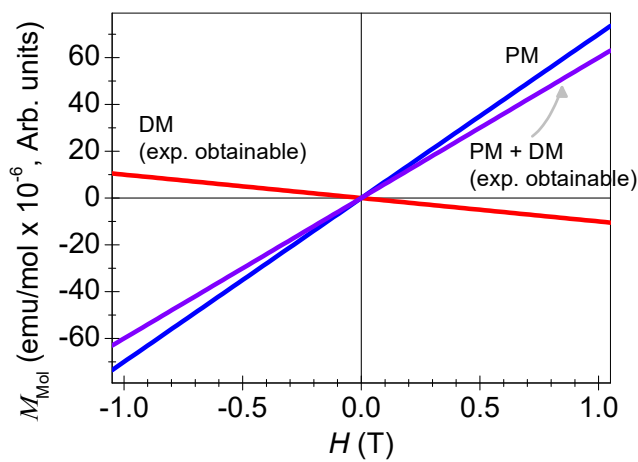

**Figure 1.** Molar magnetization $M_{Mol}$ vs. external magnetic field $H$ curves of diamagnetic (DM) and paramagnetic (PM) components of a paramagnet. The PM and DM of a paramagnet cannot be separated but this is what magnetometers measure. To obtain the PM contribution, the DM contribution is subtracted from PM + DM because they follow the additivity rule.

In the case of when the outer shell of an atom within a compound has unpaired electron(s) (and non-itinerant), then there is an additional paramagnetic contribution. The molar magnetization of these unpaired electrons $M_{Mol\_P}$ are much larger than $M_{Mol\_D}$. When the system is cooled to low enough a critical temperature, these unpaired electrons may interact with each other intermolecularly and become antiferromagnetically or ferromagnetically ordered and contribute additional magnetization.[13] For simplicity, we shall only deal with non-interacting magnetic systems, *i.e.* those that are only diamagnetic or paramagnetic, where all intermolecular electrons are treated to not interact with each other, as is the case in most MOFs under ambient conditions. Since there are no interactions, the components $M_{Mol\_D}$ and $M_{Mol\_P}$ are considered to be additive; therefore, their respective susceptibilities $\chi_{Mol\_D} + \chi_{Mol\_P} = \chi_{Mol\_D+P}$ should also be additive. For a paramagnet, only $\chi_{Mol\_D+P}$ can be measured so in order to obtain the spin contribution of the unpaired electrons, the diamagnetic contribution needs to be subtracted. Two routes are often taken to accomplish such "diamagnetic corrections". The first one is to measure an isostructural compound that is diamagnetic, which is often not feasible. The second route is to use Pascal's constants which is the summation of individual diamagnetic contributions from electrons of different core shells and bonds.[14]

Since every electron component contribution to $\chi_{Mol}$ is additive in non-interacting systems, $\chi_{Mol\_D+P}$ of two different paramagnetic compounds, A and B, should simply be the addition of $\chi_{Mol\_D+P\_A}$ and $\chi_{Mol\_D+P\_B}$. Hence, for any mixed solid $A_{1-\delta}B_\delta$ of paramagnetic compounds A and B, its $\chi_{Mol\_D+P}$ should vary linearly with respect to $\delta$. $\chi_{Mol\_D+P\_A}$ and $\chi_{Mol\_D+P\_B}$ are easily obtainable because the molar masses $m_{Mol\_A}$ and $m_{Mol\_B}$ and mass magnetizations $M_{Mass\_D+P\_A}$ and $M_{Mass\_D+P\_B}$, respectively, of A and B are measurable through the relationship:

$$\chi_{Mol\_D+P} = M_{Mass\_D+P} \cdot m_{Mol} / H \qquad (0)$$

The two measured points $(0, \chi_{Mol\_D+P\_A})$ and $(1, \chi_{Mol\_D+P\_B})$ are plotted in $\chi_{Mol\_D+P\_A}$ vs. $\delta$ space and connected by a straight line (orange dashed line in Figure 2). The equation has the general form:

$$\chi_{Mol\_D+P} = (\chi_{Mol\_D+P\_B} - \chi_{Mol\_D+P\_A}) \cdot \delta + \chi_{Mol\_D+P\_A} \qquad (1)$$

and constrains $\chi_{Mol\_D+P}$ of all mixed solids comprised of compounds A and B to have to lie on this line in the domain of $0 \leq \delta \leq 1$. Equation (1) alone is not sufficient to determine $\delta$ because $\chi_{Mol\_D+P}(\delta)$ of any mixed compound in the $A_{1-\delta}B_\delta$ system depends on $\delta$ which is unknown. More specifically, according to Equation (0), $m_{Mol}$ of a mixed sample is unknown because $\delta$ is unknown. A solution is to obtain a second equation with the same two unknown variables, $\delta$ and $\chi_{Mol\_D+P}$, but having a different slope (otherwise the two lines may be parallel and never intersect). Equation (0) has one of these unknown variables, $\chi_{Mol\_D+P}$, and $m_{Mol}$ can be parameterized to be a function of $\delta$. The molar mass of any mixed compound is actually a linear superposition of the parent compounds A and B so $m_{Mol}$ in Eq. (0) can be replaced by $m_{Mol}(\delta) = m_{Mol\_A} \cdot (1 - \delta) + m_{Mol\_B} \cdot \delta$ and $M_{Mass\_D+P} / H$ by $\chi_{Mass\_D+P}$ to yield:

$$\chi_{Mol\_D+P} = \chi_{Mass\_D+P} \cdot [m_{Mol\_A} \cdot (1 - \delta) + m_{Mol\_B} \cdot \delta] \qquad (2)$$

Here, we exploited the additivity of molar masses to obtain a second equation in the $\delta$ vs. $\chi_{Mol\_D+P}$ space with a different slope (pink dashed line in Fig. 2). The physical meaning of Eq. (2) is the mapping of $\chi_{Mol\_D+P}$ of a mixed sample with measurable $\chi_{Mass\_D+P}$ in the domain of $0 \leq \delta \leq 1$. Equation (2) has only one valid solution, the point of intersection with Eq. (1), because only at this point does $\chi_{Mol\_D+P}$ of the mixed solid lie simultaneously on Eq. (1). Hence, for any mixed sample belonging to the system $A_{1-\delta}B_\delta$, there will be a Line 1 in the form of Eq. (1) concocted by the pure compounds and each mixed sample will have its own Line 2 in the form of Eq. (2). Every mixed sample with a different δ will have a unique Line 2 because their $M_{Mass\_D+P}$ is different. Scheme 1 summarizes the procedure to obtaining δ via Equations (1) and (2).

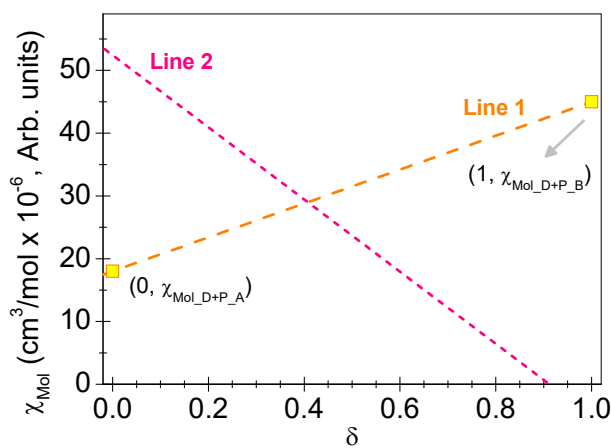

**Figure 2.** Lines 1 and 2 according to Eq. 1 and Eq. 2. Line 1 is obtained from measurements of the pure compounds and each mixed sample will have its Line 2. The intersection simultaneously yields the molar magnetic susceptibility $\chi_{Mol}$ and δ of the mixed MOF.

> **Obtaining δ and $\chi_{Mol\_D+P}$ of a mixed MOF from two pure MOFs A and B**
>
> 1. Measure mass and $\chi_{Mass\_D+P}$ of the pure MOFs A and B;
> 2. Since $m_A$ and $m_B$ are known, plot (0, $\chi_{Mol\_D+P\_A}$) and (1, $\chi_{Mol\_D+P\_B}$) and draw Line 1 to connect these two points (Equation 1);
> 3. Measure mass and $\chi_{Mass\_D+P}$ of mixed MOF of A and B and use Equation 2 to obtain a Line 2;
> 4. Point of intersection of Lines 1 & 2 will yield both δ and $\chi_{Mol\_D+P}$ of the mixed MOF.
> 5. Every mixed sample will have its own Line 2 because their $\chi_{Mass\_D+P}$ is different.

**Scheme 1.** Summarized instructions to obtaining the doping concentration δ (and therefore the mixing ratio 1–δ : δ) and molar magnetic susceptibility $\chi_{Mol\_D+P}$ of any mixed solid $A_{1-\delta}B_\delta$, where A and B are either diamagnetic or paramagnetic.

EXPERIMENTAL

MIL-101(Cr)-SO$_3$H, MIL-101(Cr)-NO$_2$, UiO-66, UiO-66-COOH, MIL-101(Cr) F Free and MIL-53(Al) were purchased from Yimobio, Inc. (China), and Leyan, Inc. (China). The compounds EuMOF (C$_{20}$H$_{17}$EuN$_6$O$_{11}$) and EuPDCA (C$_{14}$H$_{22}$EuN$_2$O$_{14}$) were synthesized according to existing literature.[15,16]

All samples were first compressed into disc-shaped pellets 3 mm in diameter. The samples were then attached onto a quartz sample holder (secured by GE Varnish) and loaded into a PPMS (Physical Properties Measurement System) manufactured by Quantum Design, Inc. The VSM (vibration sample magnetometer) option of the PPMS was used to measure the magnetization of all samples at 300 K.

RESULTS AND DISCUSSION

**Example 1: [MIL-101(Cr)-SO$_3$H]$_{(1-\delta)}$[MIL-101(Cr)-NO$_2$]$_\delta$**

One of the more popular MOFs are the MIL-101(M) series (where M = Fe, Cr, Al …).[17] Amongst them, MIL-101(Cr) is one of the most well-studied MOFs; it is extremely stable (chemically, thermally and H$_2$O retention), possesses large pore sizes, high specific surface areas and unsaturated Lewis acid sites.[18] They have applications in aqueous phase adsorption, gas sequestration & separation and catalysis.[19] As the first example, the [MIL-101(Cr)-SO$_3$H]$_{(1-\delta)}$[MIL-101(Cr)-NO$_2$]$_\delta$ system was selected because, quite frankly, they were the least expensive to acquire. Both parent compounds possess a unique element that can be used as a tracer, S and N, so ICP-MS can be employed to determine $\delta$. In comparison, the presence of paramagnetic ions makes solid-state NMR measurements hard to resolve so the samples have to undergo an elaborate process of being decomposed for solution NMR measurements.

The obtained molar magnetic susceptibilities for the two parent compounds were $\chi_{Mol\_MIL-101(Cr)-SO3H} = 10.65527(26) \times 10^{-3}$ cm$^3$/mol and $\chi_{Mol\_MIL-101(Cr)-NO2} = 31.04965(18) \times 10^{-3}$ cm$^3$/mol. Substituting the molar susceptibilities into Eq. 1 yields Line 1 for this system, which has the form $\chi_{Mol} = (\chi_{Mol\_MIL-101(Cr)-NO2} - \chi_{Mol\_MIL-101(Cr)-SO3H}) \cdot \delta + \chi_{Mol\_MIL-100(Cr)-SO3H} = [20.39438(44) \cdot \delta + 10.65527(26)] \times 10^{-3}$ cm$^3$/mol (dashed line in Fig. 3).

The molar masses of MIL-101(Cr)-SO$_3$H and MIL-101(Cr)-NO$_2$ are 957.56 and 854.35 g/mol, respectively, so Lines 2 for this system have the form $\chi_{Mol} = \chi_{Mass\_D+P} \cdot [m_{Mol\_MIL-101(Cr)-SO3H} \cdot (1 - \delta) + m_{Mol\_MIL-101(Cr)-NO2} \cdot \delta] = \chi_{Mass\_D+P} \cdot [957.56 \cdot (1 - \delta) + 854.35 \cdot \delta]$ cm$^3$/mol. The values of $\chi_{Mass\_D+P}$ for three mixed samples along with those of the parent compounds are listed in Table 1. It can therefore be seen in Fig. 3 that Lines 2 pertaining to each of the mixed samples have a slightly different slope and y-intercept, which in turn have different intersects with Line 1 yielding unique values of $\delta$.

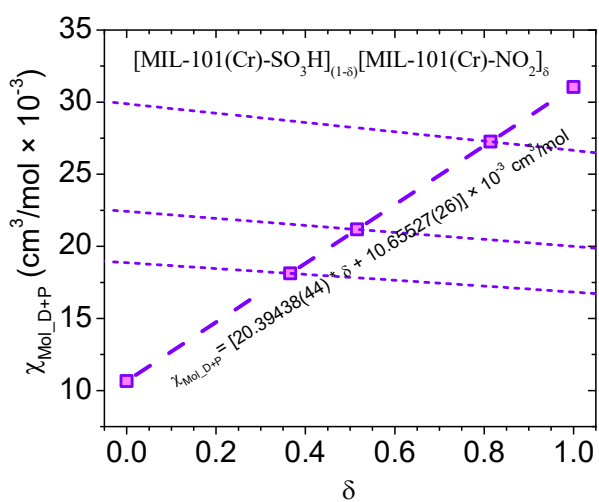

**Figure 3.** $\chi_{Mol}$ and $\delta$ of five samples belonging to the MOF system [MIL-101(Cr)-SO$_3$H]$_{(1-\delta)}$[MIL-101(Cr)-NO$_2$]$_\delta$. Line 1 (dashed line) was obtained from the pure compounds when $\delta = 0$ and 1. $\chi_{Mol}$ and $\delta$ of the mixed samples were obtained from the intersections of their Line 2 with Line 1.

The error bars are not shown because they are smaller than the data points. In the last section, error margins of $\chi_{Mol}$ and $\delta$ will be derived in detail. For the three mixed samples of the [MIL-101(Cr)-SO$_3$H]$_{(1-\delta)}$-[MIL-101(Cr)-NO$_2$]$_\delta$ system, the largest experimental error of $\delta$ was ±0.00003, meaning that $\delta$ was resolved to 33,333 parts with an average sample mass of 14.14 mg.

Table 1 also shows $\delta$ obtained from the control method $\delta_C$ of weighing the powdered samples before and after dry mixing. The small discrepancies in $\delta$ / $\delta_C$ are due to the electronic balance having an error margin, presence of impurities, and sometimes one of the parent compounds is more attracted to Zr and/or stainless steel so more of its powder is lost during the mixing and casting processes.

The Supporting Information file lists the respective diamagnetic corrections for all of the measured parent compounds and an explanation on how the effective magnetic moments $\mu_{eff}$ can be obtained. Note that diamagnetic corrections and knowledge of $\mu_{eff}$ is not necessary to determining the mixing ratios, we are just including this information for comparison purposes. In the current case, the obtained $\mu_{eff}$ of 5.17 and 8.69 $\mu_B$ for MIL-101(Cr)-SO$_3$H and MIL-101(Cr)-NO$_2$, respectively, are in agreement with existing literature concluding that the Cr$^{3+}$ ions exhibit antiferromagnetic interactions within the trimers.[9]

| $\delta_C$ (from control) | Sample mass (mg) | $\chi_{Mass\ D+P}$ (emu/g-Oe ×10$^{-6}$) | $\chi_{Mol\ D+P}$ (cm$^3$/mol ×10$^{-6}$) | $\delta$ from $\chi_{Mol\_D+P}$ | $\delta / \delta_C$ (%) |
|---|---|---|---|---|---|
| 0.00000 | 7.28 | 11.1275(3) | 10,655.27(26) | 0 | -- |
| 0.35888 | 13.54 | 19.7044(1) | 18,123.43(13) | 0.36619(2) | 2.0 |
| 0.53526 | 12.81 | 23.4113(2) | 21,171.75(21) | 0.51565(3) | – 3.7 |
| 0.79025 | 16.08 | 31.2085(1) | 27,261.30(16) | 0.81425(3) | 3.0 |
| 1.00000 | 9.51 | 36.3430(2) | 31,049.65(18) | 1.00000 | -- |

**Table 1.** Measured values of the [MIL-101(Cr)-SO$_3$H]$_{(1-\delta)}$[MIL-101(Cr)-NO$_2$]$_\delta$ system. $\delta$ was obtained from magnetic measurements and $\delta_C$ from the electronic balance to serve as a control.

**Example 2: [EuMOF]$_{(1-\delta)}$[EuPDCA]$_\delta$**

The previous example dealt with unpaired 4d electrons. However, nowadays lanthanide-based MOFs are gaining a lot of attention due to their photoluminescent properties.[20,21] Lanthanide-based MOFs possess unpaired 6f electrons, which in addition to spin, also possess an orbital angular momentum component so the proposed method also needs to be verified whether it works on mixed MOFs with lanthanide atoms. As example 2, EuMOF and Eu-PDCA were selected as we have worked with these samples in an earlier study on detection of 4-Nitrophenol.[15] These two parent compounds are comprised of only Eu, N, O, H and C so there is no unique element to serve as a tracer for ICP-MS measurements.

The observed molar magnetic susceptibilities of EuMOF and EuPDCA were 4.35738(12)×10$^{-3}$ cm$^3$/mol and 5.18724(7)×10$^{-3}$ cm$^3$/mol, respectively. After diamagnetic corrections, effective magnetic moments of 3.36 μ$_B$ and 3.63 μ$_B$ were obtained which is in good agreement with typical Eu$^{3+}$ compounds having μ$_{eff}$ in the range of 3.21-3.56 μ$_B$ subject to different ligand fields.[22]

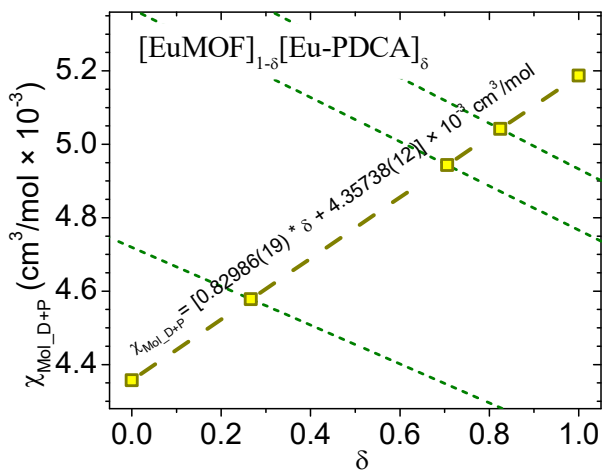

**Figure 4.** χ$_{Mol\_D+P}$ vs. δ of the [EuMOF]$_{(1–δ)}$[EuPDCA]$_δ$ system. The points of intersection are listed in Table 2.

| δ$_C$ (from control) | Sample mass (mg) | χ$_{Mass\ D+P}$ (emu/g-Oe ×10$^{-6}$) | χ$_{Mol\ D+P}$ (cm$^3$/mol ×10$^{-6}$) | δ from χ$_{Mol\_D+P}$ | δ / δ$_C$ (%) |
|---|---|---|---|---|---|
| 0.00000 | 11.51 | 6.5099(2) | 4357.38(12) | 0 | -- |
| 0.25634 | 11.53 | 7.0502(2) | 4578.23(15) | 0.26613(22) | 3.8 |
| 0.67521 | 11.07 | 8.0200(2) | 4943.23(22) | 0.70596(26) | 4.6 |
| 0.83422 | 11.92 | 8.2999(2) | 5041.77(19) | 0.82471(27) | – 1.1 |
| 1.00000 | 16.81 | 8.7284(1) | 5187.24(7) | 1.00000 | -- |

**Table 2.** Measured values of the [EuMOF]$_{(1–δ)}$[EuPDCA]$_δ$ system.

Line 1 of the [EuMOF]$_{(1–δ)}$[EuPDCA]$_δ$ system had the form: χ$_{Mol}$ = [0.82986(19) · δ + 4.35738(12)] × 10$^{-3}$ cm$^3$/mol (dashed line in Fig. 4); and Lines 2 of the three tested mixed samples had the form χ$_{Mol}$ = χ$_{Mass\_D+P}$ · [669.35 · (1 – δ) + 594.29 · δ] cm$^3$/mol (plotted as dotted lines in Fig. 4). The obtained χ$_{Mass\_D+P}$, δ and χ$_{Mol}$ are listed in Table 2. For this system, the largest obtained experimental error of δ was ±0.0003 (the average mass of the three mixed samples was 11.51 mg), a resolution larger than $10^3$. The agreement between δ and $δ_C$ indicates the applicability of the presented method on resolving the ratio of linkers between two MOFs with similar 4d (Example 1) or 6f (Example 2) electronic structure.

**Example 3: [UiO-66-COOH]$_{(1–δ)}$[UiO-66]$_δ$**

Nearly half of MOFs are diamagnetic such as NU-1000, ZIF-8, Mg-MOF-74, IRMOF-3 and UiO-67 so an example obtaining δ of mixed samples based on two diamagnetic parent MOFs is pertinent. The UiO-66 series of MOFs are $Zr^{4+}$-based so they are diamagnetic (unless host molecules with unpaired electrons are introduced). UiO-66 and UiO-66-COOH are commonly used in gas adsorption and separation and catalysis support; the latter also has usage as sensors.[23-25] A partial replacement of the covalently bonded $H^+$ in UiO-66 by $COOH^+$ renders its chemical composition hard to distinguish by ICP-MS because there are no unique elements in both compounds to serve as tracers so the system [UiO-66-COOH]$_{(1–δ)}$[UiO-66]$_δ$ was selected as Example 3. Figure 5 shows $χ_{Mol\_D+P}$ of the two parent compounds along with those of three of its mixed samples and their coinciding δ. Note that $χ_{Mol\_D+P}$ in this case is just $χ_{Mol\_D}$: there are no unpaired electrons to yield a paramagnetic contribution so $χ_{Mol\_D+P}$ is negative.

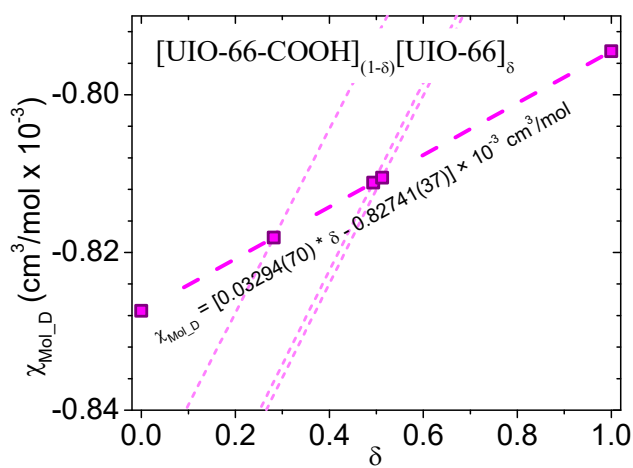

**Figure 5.** $χ_{Mol\_D+P}$ vs. δ of the [UiO-66-COOH]$_{(1–δ)}$[UiO-66]$_δ$ system. Note that these samples are diamagnetic so the values of $χ_{Mol\_D+P}$ are negative.

Line 1 of this system was $\chi_{Mol} = [0.03294(70) \cdot \delta - 0.82741(37)] \times 10^{-3}$ cm$^3$/mol (dashed line in Fig. 5); and Lines 2 of the mixed samples had the form $\chi_{Mol} = \chi_{Mass\_D} \cdot [1928.12 \cdot (1 - \delta) + 1664.06 \cdot \delta]$ cm$^3$/mol (dotted lines). Table 3 shows the measured $\chi_{Mass\_D}$ values of the mixed samples along with their δ and related parameters. The largest experimental error of δ in this system was ±0.012, the largest of all examples because the signal-to-noise ratios were smaller. This value is comparable to the ±0.01 expected value from $^1$H NMR spectroscopy, the go-to conventional method for determining molar ratios. However, the average sample mass for our method was only 10.93 mg. This result indicates that when determining the linker ratios of mixed MOFs, apart from practicality, the current method yields equally accurate results for diamagnetic mixed MOFs and even more accurate results if the mixed compounds were to be paramagnetic.

| $\delta_C$ (from control) | Sample mass (mg) | $\chi_{Mass\_D+P}$ (emu/g-Oe ×10$^{-6}$) | $\chi_{Mol\_D+P}$ (cm$^3$/mol ×10$^{-6}$) | δ from $\chi_{Mol\_D+P}$ | δ / $\delta_C$ (%) |
|---|---|---|---|---|---|
| 0.000 | 10.30 | – 0.42913(19) | – 827.41(37) | 0 | -- |
| 0.27053 | 11.90 | – 0.44135(17) | – 818.13(93) | 0.28193(1065) | 4.2 |
| 0.48463 | 10.50 | – 0.45119(19) | – 811.15(113) | 0.49349(1231) | 1.8 |
| 0.52219 | 10.40 | – 0.45210(19) | – 810.53(115) | 0.51238(1247) | – 1.9 |
| 1.000 | 10.00 | – 0.47743(20) | – 794.48(33) | 1.00000 | -- |

**Table 3.** Measured values of the [UiO-66-COOH]$_{(1–\delta)}$[UiO-66]$_\delta$ system.

**Example 4: [MIL-101(Cr) F Free]$_{(1–\delta)}$[MIL-53(Al)]$_\delta$**

Lastly, we wish to test if whether this approach is applicable to the case when one parent compound is paramagnetic and the other is diamagnetic. For the paramagnetic compound, we selected MIL-101(Cr) F Free, a commonly used MOF for gas adsorption and separation;[17,18] its measured $\chi_{Mol\_D+P}$ was 10.960000(171)×10$^{-3}$ cm$^3$/mol. After subtraction of its diamagnetic correction an effective magnetic moment of 5.21 $\mu_B$ was obtained which is in good agreement to the 5.43 $\mu_B$ result reported by Thangavel et al.[9] For the diamagnetic compound, MIL-53(Al) was chosen as it is inexpensive and thermally and chemically stable apart from its gas adsorption and storage capabilities;[26] the measured $\chi_{Mol\_D}$ was –0.094127(29)×10$^{-3}$ cm$^3$/mol. The three mixed samples for this system were intentionally selected to possess a larger fraction of the diamagnetic

MOF to test the lower limit of the proposed method since the measured magnetization would be close to zero.

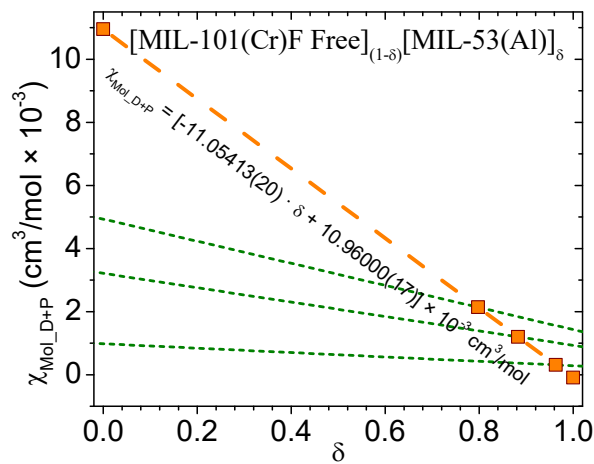

**Figure 6.** $\chi_{Mol\_D+P}$ vs. $\delta$ of the [MIL-101(Cr) F Free]$_{(1-\delta)}$[MIL-53(Al)]$_\delta$ system. MIL-53(Al) is diamagnetic so the mixed samples were deliberately prepared to possess larger amounts of MIL-53(Al) to test the accuracy of the method.

| $\delta_C$ (from control) | Sample mass (mg) | $\chi_{Mass\ D+P}$ (emu/g-Oe ×10$^{-6}$) | $\chi_{Mol\ D+P}$ (cm$^3$/mol ×10$^{-6}$) | $\delta$ from $\chi_{Mol\_D+P}$ | $\delta / \delta_C$ (%) |
|---|---|---|---|---|---|
| 0.00000 | 8.40 | 15.2723(2) | 10960.00(17) | 0 | -- |
| 0.77091 | 8.30 | 6.8762(2) | 2138.08(25) | 0.79807(5) | 3.5 |
| 0.89181 | 7.80 | 4.4783(3) | 1198.66(19) | 0.88305(5) | – 1.0 |
| 0.96877 | 11.40 | 1.3654(2) | 309.53(7) | 0.96348(4) | – 0.5 |
| 1.00000 | 14.50 | – 0.45132(1) | – 94.13(3) | 1.00000 | -- |

**Table 4.** Measured values of the [MIL-101(Cr) F Free]$_{(1-\delta)}$[MIL-53(Al)]$_\delta$ system.

Figure 6 shows Line 1 of the [MIL-101(Cr) F Free]$_{(1-\delta)}$[MIL-53(Al)]$_\delta$ system and the Lines 2 of three mixed samples. The equation of Line 1 was $\chi_{Mol\_D+P}$ = [–11.05413(20) · $\delta$ + 10.96000(17)] ×10$^{-3}$ cm$^3$/mol and the equation of Lines 2 had the form $\chi_{Mol\_D+P}$ = $\chi_{Mass\_D+P}$ · [717.37 · (1 – $\delta$) + 208.10 · $\delta$] cm$^3$/mol. The values of $\chi_{Mass\_D+P}$ and the obtained $\chi_{Mol\_D+P}$ and $\delta$ of the mixed and pure samples are listed in Table 4. An average experimental error of ±0.00005 for $\delta$ (resolution of over 2×10$^4$ parts) was obtained for this system. Since the two parent compounds have unique atoms in Cr and Al, ICP-MS applied to this particular system can resolve δ to ~$10^6$ parts. In most cases, such a high accuracy of δ is not needed. Hence, if δ only needs to be determined to an accuracy of 0.01 to 1%, then the current method has an edge in terms of practicality.

**Error analysis:**

The noise level of our magnetometer was $2\times10^{-7}$ emu at 1 T. The mass of the paramagnetic samples was typically near 10 mg, which yielded signals near ~$3\times10^{-3}$ emu so our measurements had a signal-to-noise ratio of ~$10^4$. However, this is the case if only one point was measured. In all measurements, over 900 points were obtained in between ±1 T in the span of 20 minutes so we were able to reduce the signal-to-noise ratio by an order of magnitude (conservatively). As an example, the signal-to-noise ratio for the pure MIL-101(Cr)-NO$_2$ sample was estimated as 172,811 so the obtained $\chi_{Mol}$ of 31,049.65 cm$^3$/mol×$10^{-6}$ had an error of ±$18\times10^{-8}$ cm$^3$/mol.

To obtain the errors of δ, the errors of Lines 1 and 2 have to be obtained first. Figure 7 shows Line 1 (dashed) and Line 2 (dashed) of the δ = 0.51565 sample of Example 1 along with their maximum allowed errors (solid lines). The points of intersection of the solid lines represent the expected maximum errors. The points with the largest x-component were taken as the maximum allowed error of δ. For the current example, the value of δ can be anywhere between 0.51565±3. From such, δ for this mixed sample was resolved to 33,333 parts (1/0.00003).

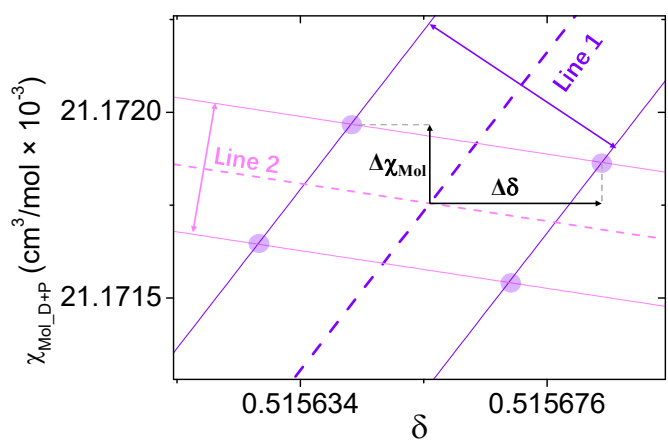

**Figure 7.** A graphical representation of how the errors of δ and $\chi_{Mol\_D+P}$ of the mixed samples were obtained. Line 1 (dashed line) along with its error margins (solid lines) is that from Example 1. Line 2 also along with its error margins is that of the mixed sample with δ = 0.51565(3) of the same system. The four dots represent the maximum allowed possible errors; the vectors with the largest *x*- and *y*-components from the intersection of the dashed lines were taken as the errors of δ and $\chi_{Mol\_D+P}$, respectively.

The noise level of the magnetometer can be further decreased by another order of magnitude if a superconducting quantum interference device (SQuID) was employed to measure the magnetic susceptibility. Along with larger sample sizes, the resolution of δ of our approach can be theoretically pushed down to 1 ppm. However, to fully appreciate such high resolutions, whether it be with our method or ICP-MS, the employed electronic balance would also have to yield a signal-to-noise ratio of $10^6$ and the samples may need to be crystallized a few times to further remove trace amounts of impurities. In our experiments, the sensitivity of our balance was only 0.01 mg and the purity of the samples varied between 95 and 98% so a difference between the theoretical δ and that of our control of ≤0.05 was reasonable. Hence, even if an electronic balance with 0.001 mg sensitivity were to be employed, the proposed method in this work of measuring the magnetic susceptibility to determining linker and node ratios of MOFs would be more practical than ICP-MS because only small amounts of samples in powdered form are needed.

Lastly, we note several potential sources that may result in erroneous results. Some MOFs can be hygroscopic so if one of the parent compounds was measured under vacuum and the mixed MOF under ambient conditions (or vice versa), then a difference in molar mass (for instance, due to adsorption of $H_2O$) will yield a false reading. It is also advised to maintain the parent and mixed MOFs not activated or all activated during the measurements; this way the molar masses do not change. Small temperature variations (for example by 1 K) may also introduce errors because the paramagnetic component is temperature-dependent. As for the possible presence of residual solvent molecules and/or extra framework paramagnetic ions in the parent compounds (which can amount to up to 4%), if the same percentage of impurities is not transferred to the mixed compound, then the accuracy will be reduced.

CONCLUSION

We provided a method to quantifying the doping ratio of mixed MOFs via measurements of the magnetic susceptibility. The main steps of the entire process are listed in Scheme 1. Four

MOF examples were worked out in detail to show how the method is practical to obtaining linker/node resolutions of ~1% in the cases when the parent compounds of the mixed samples have different organic linkers and electronic configuration. A theoretical resolution of 1 ppm can be obtained, but this would involve using expensive equipment and highly pure samples. We expect this method to be widely employed in the characterization of linker and node ratios of MOFs.

ASSOCIATED CONTENT

**Supporting Information**.

The following files are available free of charge.

Diamagnetic corrections and effective magnetic moments of the 8 parent compounds in Examples 1-4 (PDF)

AUTHOR INFORMATION

**Corresponding Author**

* fyen at hit.edu.cn or renpeng at hit.edu.cn

**Author Contributions**

The manuscript was written through contributions of all authors. All authors have given approval to the final version of the manuscript. ‡These authors contributed equally.

REFERENCES

1. Kalidindi, S. B.; Nayak, S.; Briggs, M. E.; Jansat, S.; Katsoulidis, A. P.; Miller, G. J.; Warren, J. E.; Antypov, D.; Cora, F.; Slater, B.; Prestly, M. R.; Marti-Gastaldo, C.; Rosseinsky, M. J. Chemical and Structural Stability of Zirconium-base Metal-Organic Frameworks with Large Three-dimensional Pores by Linker Engineering. *Angew. Chem. Int. Ed*. **2015**, *54*, 221-226.

2. Sun, K. K.; Li, L.; Yu, X. L.; Liu, L.; Meng, Q. T.; Wang, F.; Zhang, R. Functionalization of Mixed Ligand Metal-Organic Frameworks as the Transport Vehicles for Drugs. *J. Colloid Interf. Sci*. **2017**, *486*, 128-135.

3. Jin, J. C.; Yang, M.; Zhang, Y. L.; Dutta, A.; Xie, C. G.; Kumar, A. Integration of Mixed Ligand into a Multivariate Metal-Organic Framework for Enhanced UV-light

Photocatalytic Degradation of Rhodamine B. *J. Taiwan Inst. Chem. Eng.* **2021**, *129*, 410-417.

4. Zhao, S. S.; Zhang, H.; Wang, L.; Chen, L.; Xie, Z. Facile Preparation of a Tetraphenylethylene-Doped Metal–Organic Framework for white light-emitting diodes. *J. Mater. Chem. C* **2018**, *6*, 11701-11706.

5. Dou, S.; Song, J.; Xi, S.; Du, Y.; Wang, J.; Huang, Z. F.; Xu, Z. J.; Wang, X. Boosting Electrochemical $CO_2$ Reduction on Metal–Organic Frameworks via Ligand Doping. *Angew. Chem. Int. Ed*. **2019**, *58* 4041-4045.

6. Howarth, A. J.; Peters, A. W.; Vermeulen, N. A.; Wang, T. C.; Hupp, J. T.; Farha, O. K. Best Practices for the Synthesis, Activation, and Characterization of Metal-Organic Frameworks. *Chem. Mater.* **2017**, *29*, 26-39.

7. Zhao, M. M.; Yang, Y.; Du, N.; Zhu, Y. Y.; Ren, P.; Yen, F. Determining the Chemical Composition of Diamagnetic Mixed Solids via Measurements of the Magnetic Susceptibility. *J. Mat. Chem. C* **2024**, 12, 5877-5885.

8. Mali, G.; Mazaj, M.; Arčon, I.; Hanžel, D.; Arčon, D.; Jagličić, Z. Unraveling the Arrangement of Al and Fe within the Framework Explains the Magnetism of Mixed-Metal MIL-100(Al,Fe). *J. Phys. Chem. Lett*. **2019**, *10*, 1464-1470.

9. Thangavel, K.; Folli, A.; Ziese, M.; Hausdorf, S.; Kaskel, S.; Murphy, D. M.; Pöppl, A. EPR and SQUID Interrogations of Cr(III) Trimer Complexes in the MIL-101(Cr) and Bimetallic MIL-100(Al/Cr) MOFs. *SciPost Phys. Proc*. **2023**, *11*, 016.

10. Chavan, S. M.; Shearer, G. C.; Svelle, S.; Olsbye, U.; Bonino, F.; Ethiraj, J.; Lillerud, K. P.; Bordiga, S. Synthesis and Characterization of Amine-Functionalized Mixed-Ligand Metal-Organic Frameworks of UiO-66 Topology. *Inorg. Chem*. **2014**, *53*, 9509-9515.

11. Lammert, M.; Bernt, S.; Vermoortele, F.; De Vos, D. E.; Stock, N. Single- and Mixed-Linker Cr-MIL-101 Derivatives: A High-Throughput Investigation. *Inorg. Chem*. **2013**, 52, 8521-8528.

12. Kioka, K.; Mizutani, N.; Hosono, N.; Uemura, T. Mixed Metal–Organic Framework Stationary Phases for Liquid Chromatography. *ACS Nano* **2022**, *16*, 6771-6780.

13. Mínguez Espallargas, G.; Coronado, E. Magnetic Functionalities in MOFs: From the Framework to the Pore. *Chem. Soc. Rev*. **2018**, *47*, 533-557.

14. Bain, G. A.; Berry, J. F. Diamagnetic Corrections and Pascal's Constants. *J. Chem. Edu*. **2008**, *85*, 532-536.

15. Yang, Y.; Chen, Z.; Fu, C.; Kumar, S.; Shi, W.; Sun, F.; Yang, X.; Ren, P. Selective and Rapid Detection of 4-Nitrophenol in River and Treated Industrial Wastewater by a Luminescent Lanthanide Metal-Organic Framework Sensor. *Inorg. Chem.* **2023**, 62, 19565-19572.

16. Kong, Y. -J.; Hou, G. -Z.; Gong, Z. -N.; Zhao, F. -T.; Han, L. -J. Fluorescence Detection of Malachite Green and Cations ($Cr^{3+}$, $Fe^{3+}$ and $Cu^{2+}$) by a Europium-based Coordination Polymer. *RSC Adv*. **2022**, *12*, 8435-8442.

17. Zorainy, M. Y.; Gar Alalm, M.; Kaliaguine, S.; Boffito, D. C. Revisiting the MIL-101 Metal-Organic Framework: Design, Synthesis, Modifications, Advances and Recent Applications, *J. Mater. Chem. A* **2021**, *9*, 22159-22217.

18. Yang, C. -X.; Yan, X. -P. Metal–Organic Framework MIL-101(Cr) for High-Performance Liquid Chromatographic Separation of Substituted Aromatics. *Anal. Chem.* **2011**, *83*, 7144-7150.

19. He, R.; Fang, M.; Zhou, J.; Fei, H.; Yang, K. Multifunctional MIL-101(Cr)-$NH_2$/Expanded Graphite/Multi-Walled Carbon Nanotube/Paraffin Wax Composite Phase Change Materials with Excellent Thermal Conductivity and Highly Efficient Thermal Management for Electronic Devices. *J. Mater. Chem. C* **2023**, *11*, 11341-11352.

20. Liang, Y.; Zhu, J. -Z.; Jin, S. -Y.; Meng, Y. -R.; Li, S. -F.; Zuo, J. -L.; Zhang, G.; Su, Jian. X-ray Luminescent Metal–Organic Frameworks: Design Strategies and Functional Applications. *J. Mater. Chem. C* **2025**, *13*, 4836-4860.

21. Xie, Y.; Sun, G.; Li, J.; Sun, R.; Sun, L. $Er^{3+}$-Sensitized Upconversion/Down-Shifting Luminescence in Metal–Organic Frameworks. *J. Phys. Chem. Lett*. **2023**, *14*, 10624-10629.

22. Bronova, A.; Kannengießer, N.; Glaum, R.; Optical Spectra and Magnetic Behavior of a Wide Range of Europium(III) Oxo-Compounds: Analysis of the Ligand-Field Effects. *Inorg. Chem*. **2017**, *56*, 9235-9246.

23. Le, T. T.; Ferro-Costas, D.; Fernández-Ramos, A.; Ortuño, M. A. Combined DFT and Kinetic Monte Carlo Study of UiO-66 Catalysts for γ-Valerolactone Production. *J. Phys. Chem. C* **2024**, *128*, 1049-1057.

24. Mhatre, C. V.; Wardzala, J. J.; Oliver, M. C.; Islamov, M.; Boone, P.; Wilmer, C.; Huang, L.; Johnson, J. K. The Impact of Missing Linker Defects in UiO-66 on Adsorption and Diffusion of Isopropyl Alcohol. *J. Phys. Chem. C* **2024**, *128*, 13577-13587.

25. Zhang, N.; Yuan, L. -Y.; Guo, W. -L.; Luo, S. -Z.; Chai, Z. -F.; Shi, W. -Q. Extending the Use of Highly Porous and Functionalized MOFs to Th(IV) Capture. *ACS Appl. Mater. Interfaces* **2017**, *9*, 25216-25224.

26. Liu, J.; Li, B.; Martins, V.; Huang, Y.; Song, Y. Enhancing CO2 Adsorption in MIL-53(Al) through Pressure–Temperature Modulation: Insights from Guest–Host Interactions. *J. Phys. Chem. C* **2024**, *128*, 8007-8015.

TABLE OF CONTENTS GRAPHIC

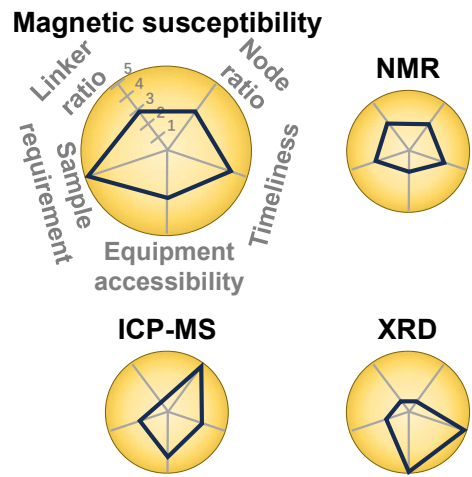